\documentclass[conference]{IEEEtran}
\IEEEoverridecommandlockouts
\usepackage{cite}
\usepackage{amsmath,amssymb,amsfonts}
\usepackage{algorithmic}
\usepackage{graphicx}
\usepackage{textcomp}
\usepackage{xcolor}
\usepackage{multirow}
\usepackage{caption}
\usepackage{subcaption}
\usepackage{comment}
\usepackage{url}
\usepackage{float}
\usepackage[T1]{fontenc}

\def\BibTeX{{\rm B\kern-.05em{\sc i\kern-.025em b}\kern-.08em
    T\kern-.1667em\lower.7ex\hbox{E}\kern-.125emX}}
\begin{document}

\title{Simulation of Entanglement Generation between Absorptive Quantum Memories\\

\thanks{This material is partly based upon work supported by the U.S. Department of Energy, Office of Science, National Quantum Information Science Research Centers. This work is also supported by Laboratory Directed Research and Development (LDRD) funding from Argonne National Laboratory, provided by the Director, Office of Science, of the U.S. Department of Energy under contract DE-AC02-06CH11357.}
}

\author{
\IEEEauthorblockN{
Allen Zang\IEEEauthorrefmark{1},
Alexander Kolar\IEEEauthorrefmark{1},
Joaquin Chung\IEEEauthorrefmark{2},
Martin Suchara\IEEEauthorrefmark{2},
Tian Zhong\IEEEauthorrefmark{1}
Rajkumar Kettimuthu\IEEEauthorrefmark{2}
}
\IEEEauthorblockA{
\IEEEauthorrefmark{1}University of Chicago, Chicago, IL, USA\\
\{yzang, atkolar, tzh\}@uchicago.edu}
\IEEEauthorblockA{
\IEEEauthorrefmark{2}Argonne National Laboratory, Lemont, IL, USA\\
\{chungmiranda, msuchara, kettimut\}@anl.gov}
}

\maketitle

\begin{abstract}
Quantum entanglement is an essential resource for quantum networks. However, the generation of entanglement between physical devices at remote network nodes is a challenging task towards practical implementation of quantum networks.
In this work, we use the open-source Simulator of QUantum Network Communication (SeQUeNCe), developed by our team, to simulate entanglement generation between two atomic frequency comb (AFC) absorptive quantum memories to be deployed on the Argonne-Chicago quantum network. We realize the representation of photonic quantum states within truncated Fock spaces in SeQUeNCe and build models for a spontaneous parametric down-conversion (SPDC) source, AFC absorptive quantum memories, and measurement devices with non-number-resolving photon detectors. Based on these developments, we observe varying fidelity with SPDC source mean photon number, and varying entanglement generation rate with both mean photon number and memory mode number. We also simulate tomographic reconstruction of the effective density matrix for the bipartite photonic states retrieved from quantum memories. Our work extends the usability of the SeQUeNCe simulator with new hardware modules and Fock state representation that will improve the simulation of near term quantum network hardware and protocols.
\end{abstract}

\section{Introduction}
\label{sec:introduction}

Quantum networks~\cite{kimble2008quantum,wehner2018quantum} promise advances in multiple fields of quantum information science including quantum communication, distributed quantum computation, and quantum metrology. Among the various resources required for practical quantum networks, quantum entanglement is one of the most important, which must be generated via non-local physical processes. In this work we simulate the generation of entanglement between two remote multi-mode atomic frequency comb (AFC) absorptive quantum memories, whose first experimental implementation has been reported in~\cite{rivera2021telecom}. 


This work makes the following contributions:
\begin{itemize}
    \item We extend the functionality of the open-source Simulator of QUantum Network Communication (SeQUeNCe)~\cite{wu2021sequence} developed by our team to support state representation of bosonic quantum systems (i.e., photonic states, within a truncated Fock space~\cite{scully_zubairy_1997}).
    \item We implement models for spontaneous parametric down conversion (SPDC) sources, AFC quantum memories, measurement devices using non-number-resolving photon detectors (with and without beamsplitter), and generalized amplitude damping channel using Kraus representation~\cite{nielsen_chuang_2010} in SeQUeNCe.
    \item We simulate entanglement generation between remote AFC quantum memories, to be deployed in a real-world quantum network prototype in the Chicago Metropolitan area. We study the entanglement fidelity and rate, while also demonstrating reconstruction of the effective density matrix of photons stored in quantum memories.
\end{itemize}

This work is organized as follows. Section~\ref{sec:background} provides background information on the simulated experiment and a brief overview of quantum network simulators. Section~\ref{sec:models} motivates the physical models of systems and processes needed for the simulated experiment. Section~\ref{sec:software_design} describes new features of the SeQUeNCe simulator developed for this work and to ease the workflow of future simulations. Section~\ref{sec:simulation_setup} describes the setup of the simulation, 
while Section~\ref{sec:simulation_results} describes and discusses the observed simulation results. Finally, we summarize our work in Section~\ref{sec:conclusion}.
\section{Background}
\label{sec:background}

\subsection{Entanglement Generation with Absorptive Memories}
Before further distribution or utilization in a quantum network, entanglement needs to be physically generated between quantum devices on remote nodes. In general, entanglement generation takes advantage of matter-photon entanglement, where two photons which are entangled with two individual quantum memories will be transmitted to an intermediate node for Bell state measurement (BSM). A successful and ideal BSM will project the joint state of two quantum memories onto a Bell state. In a recent experimental work~\cite{rivera2021telecom}, researchers achieved entanglement generation between two AFC absorptive quantum memories~\cite{Afzelius2009}. In this experiment, spontaneous parametric down conversion (SPDC) sources are the origin of matter-photon entanglement (i.e., their two output modes are entangled), where the signal mode of the output is absorbed by the AFC memory and the remaining idler mode is then entangled with the memory.

We note that AFC memories provide advantages of long storage time and multimodality. Recent research papers have presented coherent optical storage using an AFC memory protocol in rare-earth-ion doped crystals with storage time in the order of seconds~\cite{Holz_pfel_2020}, and even up to an hour~\cite{ma2021one}. Although for signals at the single-photon level the storage time is shorter, such progress still demonstrates a promising future for AFC quantum memories in long-distance quantum communication and networking. Furthermore, the physical nature of an AFC quantum memory allows for multi-mode photon storage, whereby many photons in different modes can be stored in a single memory device. This will increase the entanglement generation rate in comparison to single mode memories.

\subsection{SeQUeNCe Simulator}
In our previous work~\cite{wu2021sequence}, the Simulator of QUantum Network Communication (SeQUeNCe) was introduced, providing a framework for detailed, accurate, and highly customizable network simulations. The simulator provides a hierarchy of modules designed to replicate the potential functionality of a future quantum internet.
In this work, we focus on the two lowest modules, those for the Simulation Kernel and Hardware. The Simulation Kernel provides an interface for discrete event simulation (DES) through \texttt{Timeline} objects, as well as tracking and managing the quantum states of simulation components. The Hardware module implements numerous quantum network components, describing their behavior and modeling errors. The full simulator is open-source and available online~\cite{sequence-github}.

\subsection{Other Quantum Network Simulators}
Besides SeQUeNCe, the open-source QuISP~\cite{matsuo2019simulation} and available-upon-registration NetSquid~\cite{coopmans2021netsquid} simulators also have the goal of simulating the behavior of full-stack quantum network. QuISP pays additional attention to simulation scalability, while NetSquid has specific focus on physical layer technology, namely NV centers in diamond and atomic ensembles. 
Other simulators focused on more specialized aspects exist. For instance, QuNetSim~\cite{diadamo2021qunetsim} focuses on the upper layers of a quantum network, SQUANCH~\cite{bartlett2018distributed} has agent-based modeling which facilitates parallelization, and SimulaQron~\cite{dahlberg2018simulaqron} aims at helping quantum internet software development. Many other open-source quantum simulator projects can be found on the Quantum Open Source Foundation website~\cite{qosf}.

\section{Physical Models for Simulation}
\label{sec:models}

\subsection{AFC Quantum Memory}

The AFC memory is an essential physical component for our simulation study.
This quantum memory enables storage of photons emitted from SPDC sources and their subsequent retrieval for measurements. 
An AFC quantum memory component supports customized temporal multimode capacity which enables temporal multiplexing.
In SeQUeNCe, the AFC quantum memory interfaces with optical channel components that enable it to absorb photons and store them in predetermined temporal modes.
These temporal modes correspond to time bins that are assumed to be identical and determined by memory bandwidth.
An absorption event for the AFC quantum memory results in automatic scheduling of retrieval, where all stored photons will potentially be re-emitted in the same order as absorption according to memory efficiency.
For future simulations, this component unifies both AFC and AFC-spinwave which allows for on-demand retrieval of stored photons.
For the AFC-spinwave type, the option of photon re-emission in reversed order is available as well. The component also allows for simulation of AFC structure preparation. 
Such flexibility allows this component to be used with other quantum network protocols and application simulations.
For further details of AFC memory modeling in the simulator please refer to the source code on GitHub~\cite{sequence-github}.

\subsection{SPDC Source}
The SPDC effect enables the generation of entanglement between photonic modes.
In the experiment simulated in this work, SPDC sources are used to generate entanglement between independent AFC quantum memories.
The presence and absence of one photon in a mode ($|1\rangle$ and $|0\rangle$) defines the qubit basis (and higher excitation states thus contribute to errors), while other degrees of freedom of photons (e.g., polarization) are ignored.
The photonic state generated by the source can therefore be modeled as a two-mode squeezed vacuum (TMSV) state~\cite{lvovsky2015squeezed}
\begin{equation}
    |\psi\rangle = \frac{1}{\sqrt{\mu+1}}\sum_{n=0}^{\infty}\left(\sqrt{\frac{\mu}{\mu+1}}\right)^n|n,n\rangle
\end{equation}
where $\mu$ represents mean photon number in one output mode, and $|n,n\rangle$ represents the state where there are $n$ photons in each output mode. However, the infinite series nature of this state representation imposes the necessity for truncation. In our simulation, we allow optional truncation of Fock subspaces. With a truncated Fock space of dimension $N+1$ accounting for the normalization of quantum state, the output state is modified as
$|\psi\rangle = \sum_{n=0}^Na_n|n,n\rangle$ with amplitudes
\begin{equation}
\begin{aligned}
    a_m &= \frac{1}{\sqrt{\mu+1}}\left(\sqrt{\frac{\mu}{\mu+1}}\right)^m,\ (0\leq m<N),\\
    a_N &= \sqrt{1 - \sum_{m=0}^{N-1}\frac{\mu^m}{(\mu+1)^{m+1}}}
\end{aligned}
\end{equation}

\subsection{Quantum Measurement and Measurement Devices}
According to the postulates of quantum mechanics, measurements are described by a set of measurement operators $\{M_m\}$ where the measurement outcome $m$ occurs with a probability $p(m)=\langle\psi|M_m^{\dagger}M_m|\psi\rangle = \mathrm{tr}(M_m\rho M_m^{\dagger})$. 
The set of operators $\{M_m\}$ thus must satisfy $\sum_mM_m^{\dagger}M_m=I$ (to ensure the measurement outcome probability distribution is normalized) and the post-measurement state is given by 
\begin{equation}
    |\psi\rangle_m = \frac{M_m|\psi\rangle}{\sqrt{\langle\psi|M_m^{\dagger}M_m|\psi\rangle}},\ \rho_m = \frac{M_m\rho M_m^{\dagger}}{\mathrm{tr}(M_m\rho M_m^{\dagger})}
\end{equation}

A more general framework of quantum measurement, the positive operator-valued measure (POVM), focuses on statistics of detection results. The probability of getting a certain measurement result $n$ is given by
\begin{equation}
    P_n = \mathrm{Tr}(\rho\Pi_n)
\end{equation}
where $\rho$ is the density matrix of quantum state to be measured and $\Pi_n$ is the POVM operator corresponding to measurement result $n$ which also satisfies the normalization requirement $\sum_n\Pi_n=I$. There is a degree of freedom in deciding the post-measurement state of a POVM, which actually depends on the physical realization of such POVM.
For simplicity, in this work we define the measurement operators $M_m\equiv\sqrt{\Pi_m}$ which automatically satisfy the normalization condition.

In our simulation, measurements are done by photon detectors that we assume are non-number-resolving due to technological limitations. Their corresponding POVM operators could be constructed as~\cite{Chou2005}
\begin{gather}
\label{eqn:povm}
    \Pi_1 = \sum_{n=1}^{\infty} (-1)^{n+1}\frac{a^{\dagger n}a^n}{n!} = I - |0\rangle\langle 0|,\\ \Pi_0 = I - \Pi_1 = |0\rangle\langle 0|
\end{gather}
where $a^{(\dagger)}$ is the annihilation (creation) operator of the photonic mode to be measured, $I$ is the identity operator, and $|0\rangle$ is the Fock ground state.
If we want to consider the detector inefficiency, we replace the ``bare'' operators with new ones modified by a coefficient~\cite{Chou2005}
\begin{equation}
    a\rightarrow\sqrt{\eta}\,a
\end{equation}
where $0\leq\eta\leq 1$ is the detector efficiency.
When working in truncated Fock space for every subsystem, it is desired that states will never surpass the truncation. Therefore, we implement the effect of beamsplitters for measurement devices by transforming the POVM operators rather than the input quantum states. This is because the output state of a beamsplitter may exhibit a photon bunching effect that will bring the state beyond the original truncated space. 

\subsection{Generalized Amplitude Damping Channel}
Photon loss is ubiquitous in photonic experiments, arising from optical fiber transmission or photon absorption and retrieval.
The effect of photon loss on photonic states can be captured by the generalized amplitude damping channel~\cite{chuang1997bosonic}. This quantum channel can be expressed explicitly under Kraus (operator-sum) representation ${\cal E}(\rho)=\sum_kE_k\rho E_k^{\dagger}$ where $\sum_{k}E_k^{\dagger}E_k=I$.
The Kraus operators are
\begin{equation}
    E_k=\sum_{n\geq k}\sqrt{\binom{n}{k}}\sqrt{(1-\gamma)^{n-k}\gamma^k}\,|n-k\rangle\langle n|
\end{equation}
where $\gamma$ is the loss probability for a single photon, and $k$ can be interpreted as the number of lost photons.
The above series has again infinitely many terms; in simulation we restrict the upper limit of summation index to be $N$ for a truncated Fock space of dimension $N+1$, and it is straightforward to verify that such truncated Kraus operators still satisfy the normalization condition $\sum_{k=0}^NE_k^{\dagger}E_k = \sum_{n=0}^N|n\rangle\langle n|$.

\section{Software Design}
\label{sec:software_design}

\subsection{Modularized Hardware Configuration}
To ease the development of this simulation study and of future independent studies, the hardware module of SeQUeNCe has been modified to provide more intuitive and easily configurable interfaces. The basic functionality of the hardware module, along with several example hardware components, is discussed in our previous work~\cite{wu2021sequence}.

\subsubsection{Modifications to the Entity Class}
All physical simulation components (those from the hardware module) derive from SeQUeNCe's \texttt{Entity} class, which requires components to define a string \texttt{name} and a reference to a \texttt{Timeline} object used to execute simulation events. For this work, we have additionally required all simulation elements in a local area (including optical components and nodes, but not channels) to implement a \texttt{get} method along with a list of other local entities \texttt{receivers}. The \texttt{get} method allows any local object to receive a \texttt{photon} during the simulation, and if the \texttt{photon} must be forwarded, one of the entities in the \texttt{receivers} list is chosen. If the chosen receiver is a node, the node may run code to determine a channel to send the \texttt{photon} through (if necessary). This change serves to simplify the method of adding new components to the simulator and allows for a more streamlined process of configuring local hardware.

Nodes additionally maintain a \texttt{components} dictionary that stores all local components mapped by component \texttt{name}. Rather than relying on discrete node attributes that reference hardware elements (as in previous versions), protocols may now function on a variety of node types provided they know the name of the required hardware components. This allows for more flexibility in selecting node hardware elements without changing fundamental processes and code on a node.

\subsubsection{Modifications to the Photon Class}
\label{sec:photon}
At the heart of physical simulation in SeQUeNCe is the \texttt{Photon} class. This class simulates individual photons, which perform a range of functions from storing a local quantum state to heralding the state of quantum memories. Due to this wide and increasing range of uses, the class has been rewritten to handle multiple usage options and facilitate further development.

The \texttt{quantum\char`_state} field of \texttt{photon} has been modified to allow for two separate usage paradigms. For simulations relying on minimal photon entanglement and short-lived photons such as quantum key distribution or teleportation between adjacent nodes, the quantum state of the photon is usually small in size and created and destroyed within a short time span. For this usage, the \texttt{quantum\char`_state} field directly stores a local object that is destroyed with the photon. For more complex simulations, or those involving photon storage, the quantum state is stored in the quantum state manager (QSM) of the simulation \texttt{Timeline}. This allows many photons to easily share a single composite state (usually entangled), provides more robust error modeling, and allows for parallel network simulation. The \texttt{quantum\char`_state} field then stores a key to the shared quantum state within the manager. 

\subsection{Quantum State Modeling and Storage}
As mentioned in~\ref{sec:photon}, quantum states are stored in the simulation QSM. To facilitate more realistic simulation, we have extended the existing SeQUeNCe storage formalisms to support storage of bosonic Fock state density matrices. Single-partite states are stored directly as $(N+1) \times (N+1)$ matrices, where $N$ is the Fock space truncation parameter that may be set by the user. The QSM provides external interfaces for generating simple operators, applying operators to states, applying generalized amplitude damping channel to state, and state measurement, allowing hardware models to be constructed with minimal knowledge of the underlying state representation.

\subsection{Hardware Components}
After standardizing the usage and structure of the SeQUeNCe hardware module, it was necessary to add new hardware components to reproduce the experiment described in~\cite{rivera2021telecom} using models from Section~\ref{sec:models}. These components provide SeQUeNCe with a greater range of functionality for future network simulations. 

\section{Simulation Setup}
\label{sec:simulation_setup}

\subsection{Argonne-Chicago Quantum Network}


The simulation setup is inspired by the Argonne-Chicago Quantum Network and the Quantum Link Lab facilities. The quantum network facilities consist of four geographically separated network nodes and four multistrand optical fiber links totaling over 452 km of fiber capable of carrying classical and quantum network information. The network nodes are the Argonne Central node (ANL) located on Argonne's main campus, Argonne-in-Chicago node at Harper Court (HC), the University of Chicago node at the Eckhardt Research Center (ERC), and the Fermilab (FNAL) node. This network is designed for demonstrating entanglement distribution between multiple users.

\subsection{Optical Hardware Configuration}

\begin{figure}[htp]
\centering
\includegraphics[width=0.8\columnwidth]{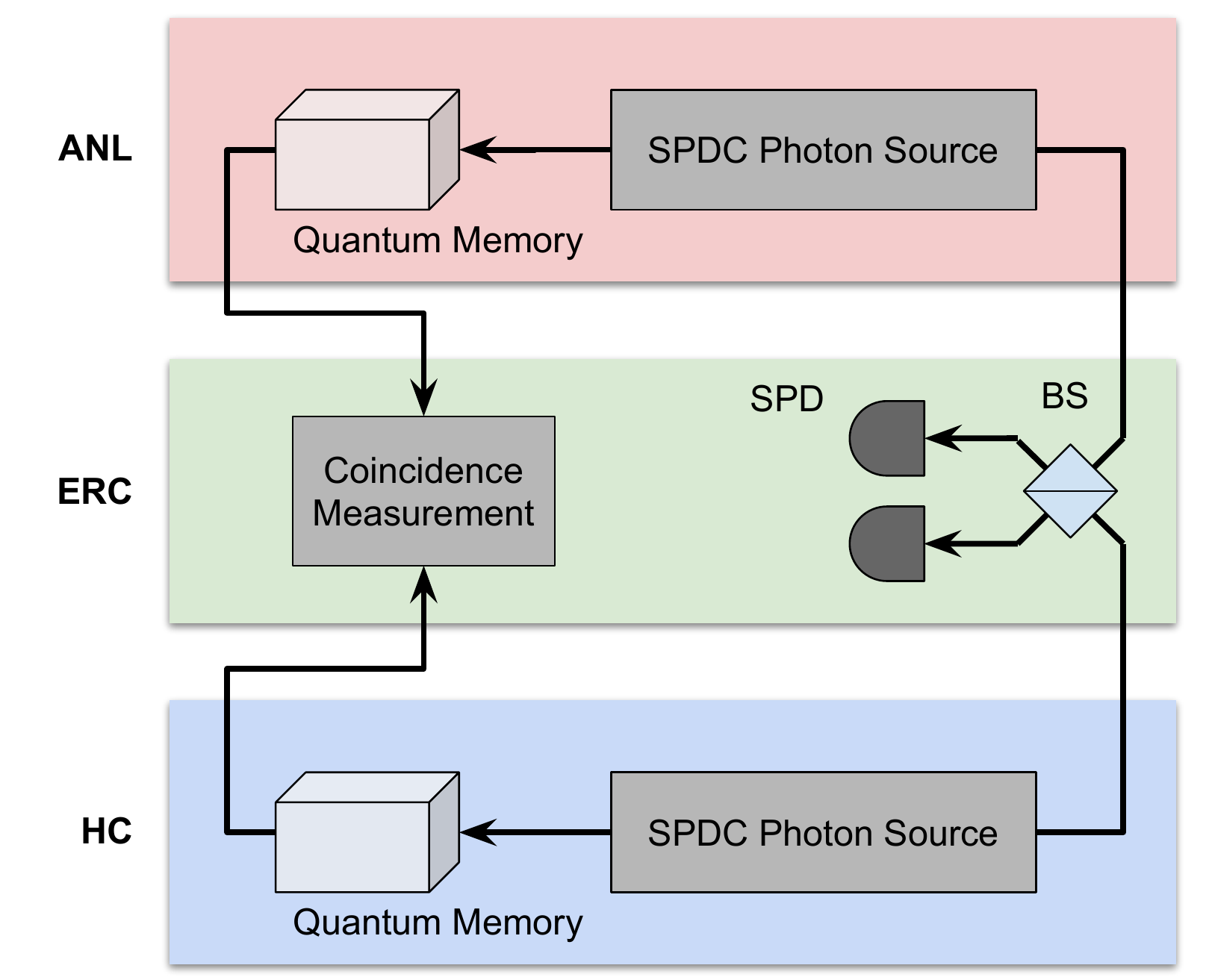}
\caption{Hardware setup used in the simulation. Entanglement between photons stored on the ANL and HC nodes is established at the ERC node, where the entanglement is also measured and quantified.}
\label{fig:hardware}
\end{figure}

The described network configuration was simulated using models of the optical hardware elements in~\cite{rivera2021telecom}, with ANL and HC nodes both holding a quantum memory and an SPDC source, and ERC holding measurement devices. To facilitate timing synchronization, we assume that the optical fiber link between ERC and HC is extended to the same length as the fiber link between ERC and ANL, which is feasible experimentally.
Each SPDC source is connected to the AFC quantum memory of its respective node and to a beamsplitter (BS) and single photon detectors (SPD) on the measurement node (ERC). The BS and SPDs are integrated into a single Quantum State Detector (QSD) device. The whole setup is shown in Figure \ref{fig:hardware}. 

During simulation, idler photons from each SPDC source are sent to the intermediate measurement node, where successful Bell state measurement (BSM) with only one photon detection click ideally heralds the entanglement state
\begin{equation}\label{eq:mem_state}
    |\psi\rangle = \frac{1}{\sqrt{2}}\left(|0\rangle_A|1\rangle_B \pm e^{i\Delta\varphi}|1\rangle_A|0\rangle_B\right)
\end{equation}
between the two quantum memories, where the relative sign depends on the photon detector that clicks. Here, the states $|0\rangle_N$ and $|1\rangle_N$ represent 0 or 1 collective atomic excitation, respectively, within the quantum memory $N$. $\Delta\varphi$ denotes the relative phase difference between the two idler paths; for simplicity we assume $\Delta\varphi=0$ in our simulations. The signal photons are absorbed by the quantum memories on the ANL and HC nodes, and are measured after automatic re-emission by other devices on the intermediate ERC node.


\subsection{Quantum State Recovery}
\label{sec:recover}
After establishing entanglement between the two AFC quantum memories, we characterized the generated quantum state. We first directly access and evaluate this state as stored within the simulation; additionally, we simulate a tomography experiment using a combination of diagonal and coherence measurements as described in~\cite{rivera2021telecom,Chou2005}. The effective quantum state of the memory modes can be represented as a $4\times 4$ density matrix if we restrict ourselves within a truncated Fock space where every mode has at most one excitation. In the basis $|m\rangle_{ANL}|n\rangle_{HC}$, where ${m,n} = {0,1}$ represents the number of photons in each mode, we have
\begin{equation}
    \Tilde{\rho} = \frac{1}{\Tilde{P}}\begin{bmatrix}
    p_{00} & 0 & 0 & 0 \\
    0 & p_{01} & d & 0 \\
    0 & d^* & p_{10} & 0 \\
    0 & 0 & 0 & p_{11}
    \end{bmatrix}
\end{equation}
where $p_{ij}$ will be recovered from the probability of measuring $i$ ($j$) photons in the ANL (HC) mode by taking inefficiencies into consideration according to discussion in Methods section of~\cite{rivera2021telecom}. $\Tilde{P} = p_{00} + p_{01} + p_{10} + p_{11}$ is a normalization factor introduced 
so that $\Tilde{\rho}$ has trace one.

\begin{figure}[htp]
\centering
\includegraphics[width=\columnwidth]{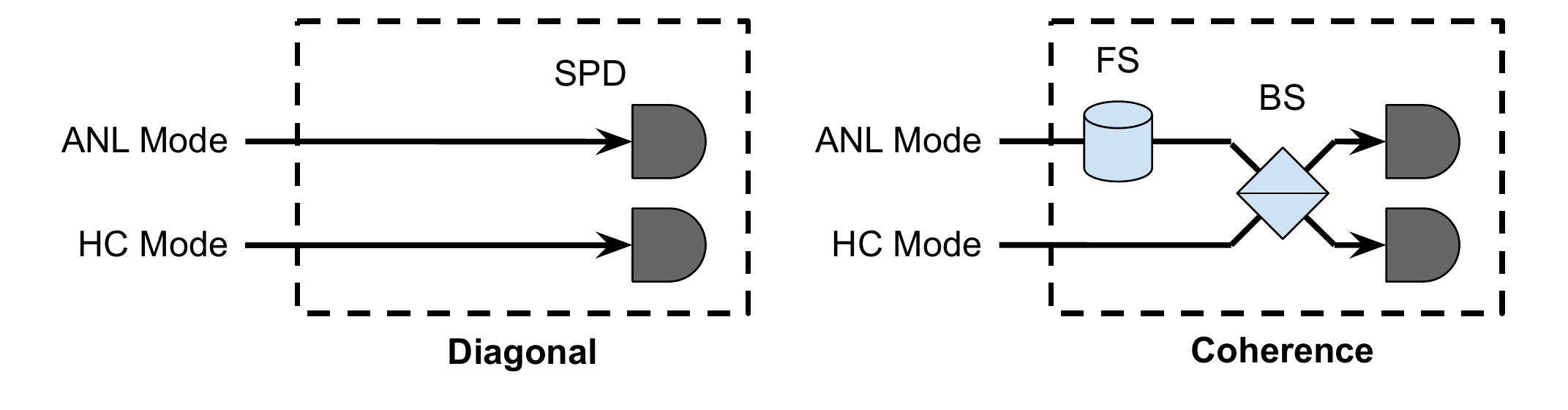}
\caption{Hardware setup used in the simulation to measure the entangled memory state. The configuration in the ``Diagonal'' box is for measuring diagonal elements, while the ``Coherence'' configuration is for measuring the off-diagonal element $d$.}
\label{fig:hardware_meas}
\end{figure}

To determine the diagonal matrix elements, two SPDs were used to detect photons in each mode directly. The amplitude of the off-diagonal amplitude $d$ is revealed by introducing a BS for mode interference, coupled with a fiber stretcher (FS, also incorporated into a QSD) to control the relative phase of photonic modes retrieved from the two memories. We then have that $|d| \approx V(p_{10}+p_{01})/2$, where $V$ is the visibility of the resulting interference pattern. The hardware configuration used to perform these measurements is shown in Figure~\ref{fig:hardware_meas}.

\section{Simulation Results}
\label{sec:simulation_results}


\subsection{Effective Entanglement Fidelity and Generation Rate}
\label{sec:fidelity_rate}
In this section, we present the directly calculated fidelity of the generated entangled state, which is stored in AFC quantum memories conditioned on successful BSM. We also show the calculation of the entanglement generation rate to demonstrate the expected performance of entanglement generation under varying experimental parameters.

To calculate the fidelity, we directly access the joint state of two quantum memories' storage modes if a heralding signal from the BSM node is received, which we denote as $\rho$. We note that due to inevitable memory absorption loss, the post-BSM state of memories will contain a significant portion of the $|00\rangle\langle 00|$ component, which is absent in the ideal Bell state for absence-presence encoding. To deal with this photon loss effect, we can introduce an effective density matrix by restricting the matrix element of $|00\rangle\langle 00|$ to zero, and then normalizing the modified density matrix. We denote the new effective density matrix as $\Tilde{\rho}$. This can be intuitively understood by considering that the $|00\rangle\langle 00|$ term will not generate detectable effect if $\rho$ is utilized in the future (e.g., for entanglement purification or teleportation) and therefore the effect of this term can be eliminated via post-selection. The fidelity can be calculated in a straightforward manner by $F = \mathrm{tr}(\Psi_{\pm}\Tilde{\rho})$ given that reference Bell state $\Psi_{\pm}$ is pure.

\begin{figure}[h]
    \centering
    \includegraphics[width=0.9\columnwidth]{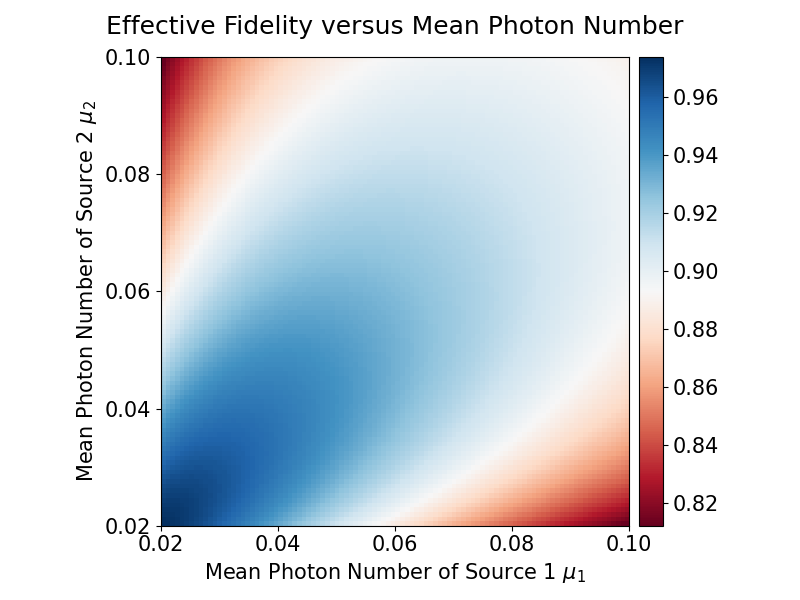}
    \caption{Effective fidelity of generated entanglement as function of mean photon numbers of two SPDC sources.}
    \label{fig:fidelity}
\end{figure}

In Figure~\ref{fig:fidelity}, we demonstrate how fidelity of the generated entanglement between two AFC quantum memories varies with different mean photon numbers in the output ports of the two SPDC sources $\mu_{1,2}$. We considered a symmetric experiment setup with identical AFC memory absorption efficiencies $\eta_\mathrm{AFC} = 0.35$, identical fiber lengths $\ell = 20\,\mathrm{km}$ and attenuation $\alpha = 0.2\,\mathrm{dB/km}$, and identical detector efficiency $\eta_\mathrm{d} = 0.6$ for the two detectors performing BSM. The figure is obviously symmetric along the $\mu_1=\mu_2$ axis as a result of our symmetric choice of simulation parameters for this calculation. Another feature is that when $\mu_1=\mu_2$, the effective fidelity is the highest, while lower $\mu_{1,2}$ additionally offers higher effective fidelity. The first feature can be understood through simplified analysis with a pure initial state
\begin{equation}
    |\psi\rangle = (c_{10}|00\rangle_{01}+e^{i\phi_1}c_{11}|11\rangle_{01})\otimes(c_{20}|00\rangle_{23}+e^{i\phi_2}c_{21}|11\rangle_{23})
\end{equation}
where 0 and 3 are memory modes while 1 and 2 are photonic modes to be measured. Straightforward derivation gives that an ideally successful BSM will project the joint state of the memories onto
\begin{equation}
    |\psi\rangle_\mathrm{m} \propto e^{i(\phi_1-\phi_2)}c_{11}c_{20}|10\rangle_{03} \pm c_{10}c_{21}|01\rangle_{03}
\end{equation}
With a symmetric setup, the decays of the $|11\rangle$ term are identical for both optical paths, thus the changes in $c_{11}$ and $c_{10}$ are equivalent. Generated states with higher fidelity have more similar values for $c_{11}c_{20}$ and $c_{10}c_{21}$, thus it is intuitive that if the initial coefficients $c_{10}=c_{20}$ and $c_{11}=c_{21}$ the fidelity will be optimal. Additionally for the second feature, consider real SPDC sources whose output states contain $|22\rangle$ and even higher order components. For lower mean photon number, the amplitudes of the higher order components which represent errors are also lower, therefore the fidelity is higher.

For the entanglement generation rate, we consider a single entanglement generation cycle where two trains of photons are emitted and BSM results will be classically communicated back to the memory nodes. The total duration of this cycle is 
\begin{equation}
    T = \frac{M}{f} + \tau_\mathrm{ph} + \tau_\mathrm{c}
\end{equation}
where $M$ is the number of emitted pulses from one SPDC source during one cycle, which is assumed to be the mode number of AFC memories in the simulation, $f$ is the frequency for SPDC source to emit photons ($1/f$ thus defines the characteristic time scale for a single wavepacket), $\tau_\mathrm{ph}$ is the fiber delay of transmitted photons, and $\tau_\mathrm{c}$ is the classical communication delay. Given the average number of entangled pairs successfully generated in one cycle $N = Mp_\mathrm{h}$, where $p_\mathrm{h}$ is the heralding probability, the entanglement generation rate is given by
\begin{equation}
    R = \frac{N}{T} = \frac{Mp_\mathrm{h}}{M/f + \tau_\mathrm{ph} + \tau_\mathrm{c}}
\end{equation}
\begin{figure}[h]
    \centering
    \includegraphics[width=0.9\columnwidth]{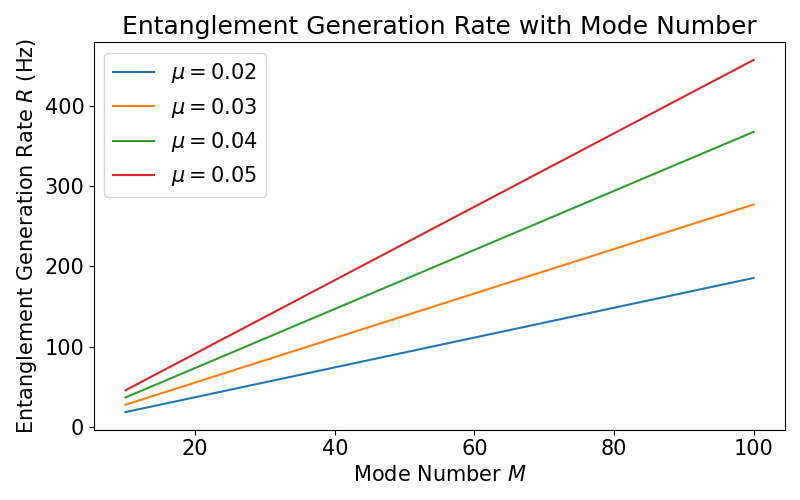}
    \caption{Entanglement generation rate as function of memory mode number for different SPDC output mean photon numbers.}
    \label{fig:rate}
\end{figure}

The temporal multiplexing advantage from memory multimodality is shown in Figure~\ref{fig:rate}, with simulation parameters unchanged from those for the effective fidelity of entanglement. The near-linear relation between generation rate and memory mode number is observed as expected~\cite{simon2007quantum}, along with increased rate for higher mean photon number. Note that the effective fidelity of the generated entanglement is lower for higher mean photon number, representing a trade-off between rate and fidelity. The rate at few hundred Hz is also compatible with the order of magnitude reported in the experiment~\cite{rivera2021telecom}.

\subsection{Density Matrix Reconstruction}

\begin{figure}[h]
    \centering
    \begin{subfigure}[b]{0.75\columnwidth}
        \centering
        \includegraphics[width=\textwidth]{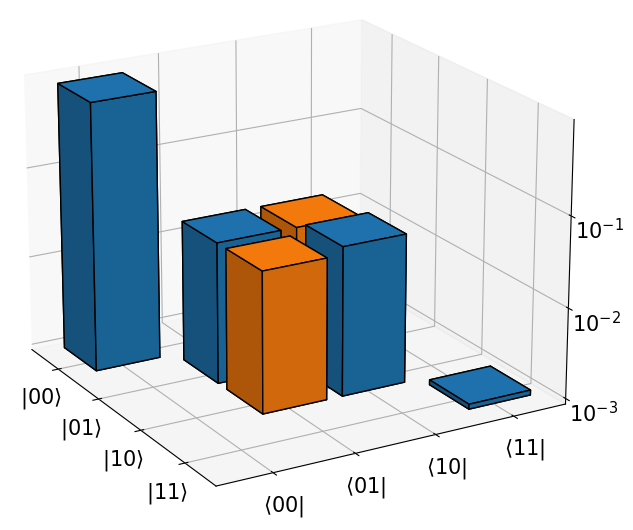}
        \caption{Reconstructed density matrix. Note the logarithmic scale for the height.}
        \label{fig:density}
    \end{subfigure}
    \hfill
    \begin{subfigure}[b]{0.9\columnwidth}
        \centering
        \includegraphics[width=\textwidth]{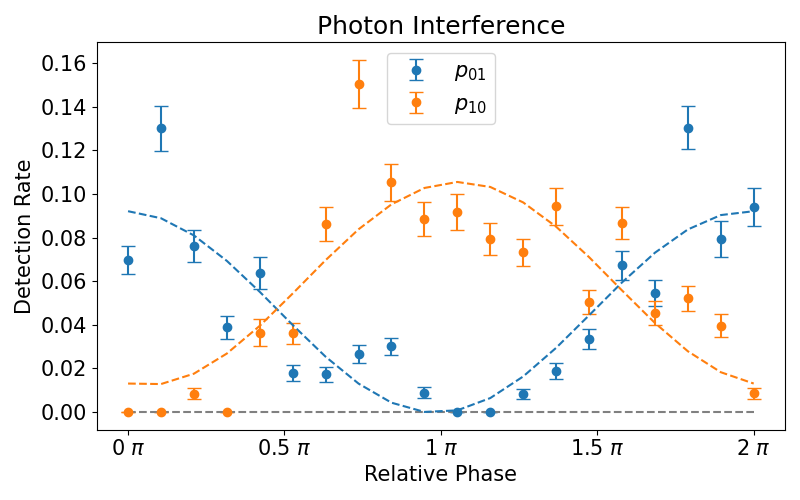}
        \caption{Photon detection probabilities at each BS output detector, conditioned on successful BSM, along with fitted sinusoid. Error bars represent one standard deviation.}
        \label{fig:interference}
    \end{subfigure}
    \caption{Simulated entanglement characterization.}
    \label{fig:measurement}
\end{figure}

In this section, we characterize the generated entangled state using a simulated state tomography experiment as described in~\ref{sec:recover}. We use realistic parameters for the setup, including SPDC photon number $\mu_{1,2} = 0.1$, detector efficiency $\eta_d = 0.6$, detector dark count rate $D = 150\,\mathrm{Hz}$, fiber length $\ell = 20\,\mathrm{km}$, fiber attenuation $\alpha=0.2\,\text{dB/km}$, and memory efficiency $\eta_\mathrm{AFC} = 0.35$. The results are shown in Figure~\ref{fig:measurement}.

The residual $p_{00}$ value is primarily produced by detector inefficiency and fiber/memory loss, while the $p_{11}$ value is produced by detector dark counts and BSM inefficiencies.

Figure~\ref{fig:interference} showcases the interference experiment used to measure the off-diagonal elements in the density matrix. The detection rate at each output mode of the Coherence detector configuration (see Figure~\ref{fig:hardware_meas}) was recorded as a function of the relative phase between the input modes, conditioned on a detection in a \textit{single} output mode during BSM measurement. To calculate the visibility, a sinusoid was fit to the detection rate for each detector. An average visibility of 89.8\% was observed for the two detectors, giving an off-diagonal magnitude $|d| = 0.0361$ plotted in Figure~\ref{fig:density}. These visibilities are high compared to similar experimental setups, arising due to the absence of phase error models in the current simulation.

\section{Conclusion}
\label{sec:conclusion}

In this work, we have extended SeQUeNCe to provide for a more accurate physical simulation in line with other lower-layer network simulators. This includes streamlining of hardware models to better facilitate addition and modification of components, as well as implementing a more robust formalism for state storage and modification. With these modifications, we have performed a simulation of a realistic experiment of entanglement generation between two AFC quantum memories. We demonstrate the effective fidelity of the generated quantum state and entanglement generation rate as functions of tunable experimental parameters, and simulate state tomography assuming no direct access to the stored quantum state by observers. Noting the realism of these experiments, and citing the open-source nature of SeQUeNCe, we provide an efficient and accurate means to 
evaluate design choices for future quantum networks. We leave the exploration of broader experimental parameter space and practical application of generated entanglement for future work.


\newpage

\bibliographystyle{./references/IEEEtran}
\bibliography{./references/references}


\end{document}